# Non-local Aharonov-Bohm conductance oscillations in an asymmetric quantum ring

S.S. Buchholz[a)], S.F. Fischer, and U. Kunze,
*Werkstoffe und Nanoelektronik, Ruhr-Universität Bochum, D-44780 Bochum, Germany*

D. Reuter and A. D. Wieck
*Angewandte Festkörperphysik, Ruhr-Universität Bochum, D-44780 Bochum, Germany*

We investigate ballistic transport and quantum interference in a nanoscale quantum wire loop fabricated as a GaAs/AlGaAs field-effect heterostructure. Four-terminal measurements of current and voltage characteristics as a function of top gate voltages show negative bend resistance as a clear signature of ballistic transport. In perpendicular magnetic fields phase-coherent transport leads to Aharonov-Bohm (AB) conductance oscillations which show *equal* amplitudes in the local and the non-local measurement at a temperature of 1.5 K and above. We attribute this novel observation to the symmetry of the orthogonal cross junctions connecting the four quantum wire leads with the asymmetric quantum wire ring.

The wave nature of charges becomes particularly evident in interference phenomena. 1D quantum ring structures have been applied to measure interference effects of coherent electron flow[1-3] and to investigate phase coherence in the electron transport through embedded quantum dots.[4-6] Controlled collimated coherent electron beams[7] of selected wave vectors are of topical interest in nanoelectronic and spintronic device research. In particular, various theoretical proposals exist for spin filtering, spin manipulation and spin detection in 1D channels and quantum wire ring systems.[8-12] Thus to-date, fundamental aspects of 1D quantum transport in complex quantum wire systems, such as quantum rings with multiple quantum wire leads require experimental exploration.

Here, we investigate firstly the ballistic transport in four quantum wire leads that feed an utmost *asymmetric* quantum wire ring. The asymmetry of the quantum wire ring allows to investigate the transmission coefficients for the two unequal paths. Based on the four quantum wire leads quantum interference phenomena in a *local* and *non-local* configuration are measured. We demonstrate that the observed AB conductance oscillations have equal amplitudes in both measurement configurations. The high quality quantum wire structure shows AB conductance oscillations at temperatures up to $T \sim 2$ K. This is relevant for the implementation of theoretical proposals based on complex quantum wire ring systems, such as spin interferometers.[8-12]

Quantum interference in electron transport through nanoscale ring structures can be studied by the magnetic Aharonov-Bohm (AB) effect.[1] The total phase change in the electron wave function is the sum of orbital and AB phase, and the total transmission oscillates with varying enclosed magnetic flux $\phi$ with a period of $\phi_0 = h/e$. In AB interference experiments the current is usually passed through the ring, and the voltage is measured in a two-terminal[13] or a local four-terminal setup[14]. Kobayashi and co-workers[15] performed experiments in a non-local setup of a *symmetric* quantum ring and showed that electron wave function decoherence in a symmetric AB ring is probe-configuration dependent.[15] In the non-local measurement setup the decoherence is strongly reduced compared to the local setup.[15,16] Due to the symmetry of the ring under investigation in,[6,15] inertial ballistic characteristics could not be detected: Collimation leads to an equal injection of ballistic electrons in both arms of the ring. Here, we chose an asymmetric quantum ring in order to investigate the transmission in the strongly differing paths of ring structure. Remote bend resistance measurements and quantum interference effects are performed and discussed with relation to the geometry of the nanoscale cross junctions which attach the leads to the ring.

Our quantum wire samples were fabricated from a modulation doped AlGaAs/GaAs heterostructure grown by molecular-beam expitaxy. A two-dimensional electron gas is situated at the heterojunction 55 nm below the surface. The 2DEG has an electron density of $3.1 \times 10^{11}$ cm$^{-2}$ and a mobility of $1 \times 10^{6}$ cm$^2$/Vs, measured in the dark at $T = 4.2$ K. This provides a mean free path of about $l_e \sim 9.5$ µm. The nanoscale quantum channels and the macroscopic leads and electron reservoirs were defined by a mix-and-match process combining low-energy electron beam lithography (2 kV, area dose 800 µC/cm$^2$) and standard optical lithography. Special care was taken to minimize the proximity effect in order to avoid widening of nanoscale corners at the quantum wire cross junctions. Subsequently, the complete device pattern was transferred into the heterostructure in one processing-step by anisotropic wet-chemical etching with an aqueous-citric acid solution ($C_6H_8O_7$:$H_2O_2$:$H_2O$; 5:1:20, etch duration of 52 s). The etch depth was determined by atomic force microscopy and amounts to 40 nm. AuGeNi ohmic contacts and a global Au top-gate were fabricated by photo-

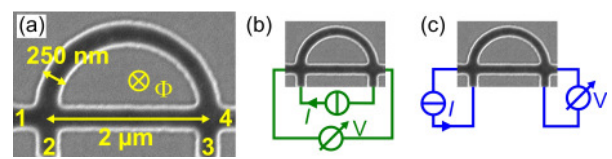

Fig. 1. (a) Scanning electron micrograph of the etched ring structure before the deposition of an Au gate electrode. (b) Local four-terminal measurement setup with an input current from lead 2 to lead 3. (c) Non-local four-terminal setup with an input current from lead 2 to lead 1.

---
[a)] sven.buchholz@rub.de

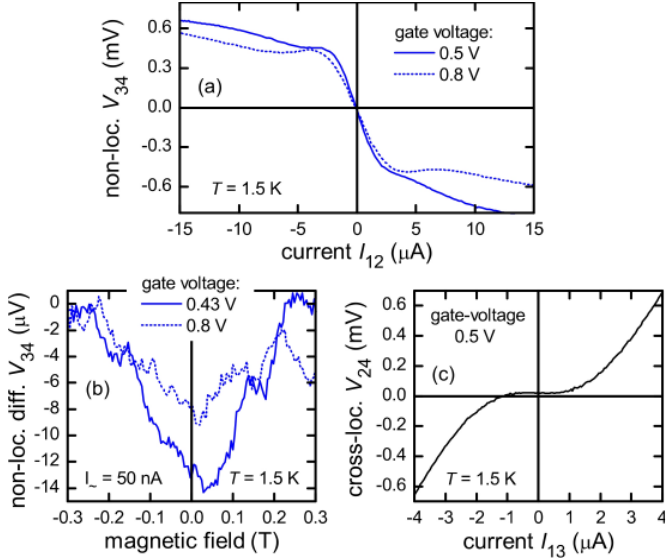

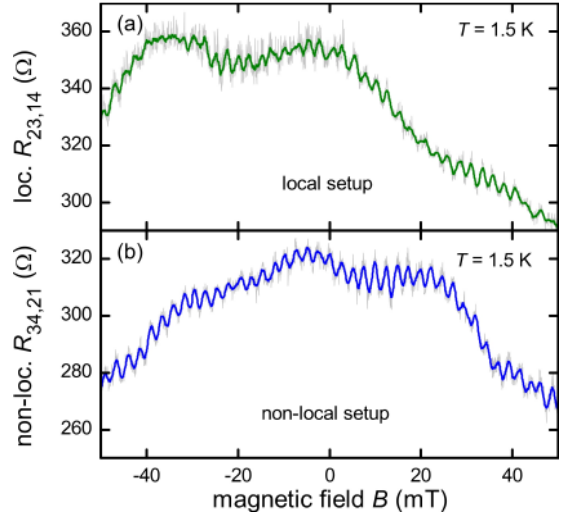

Fig. 2. (a) Bend resistance characteristic: Non-local voltage versus DC input current. For $-2\mu A < I_{12} < 2\mu A$ a strong negative resistance is observed. (b) Differential bend resistance versus a perpendicular applied magnetic field. (c) Local four-terminal voltage $V_{24}$ versus DC input current $I_{13}$ measured in the cross-local configuration.

Fig. 3. (a) Local differential resistance $R_{23,14}$ and (b) non-local $R_{34,21}$ characteristic in a perpendicular magnetic field.

lithography, thermal-evaporation and subsequent lift-off processing. The AuGeNi contacts were alloyed by rapid thermal annealing (25 s at ~380°C).

Fig. 1 shows a scanning electron microscope image of the ring structure before the deposition of the top-gate electrode. The ring structure consists of two equally 250 nm-wide electron waveguides which intersect twice in orthogonal cross junctions. The straight and the bent waveguides have a total length of 2.8 µm and 4.1 µm, respectively, and are attached to the reservoirs adiabatically. The distance between the intersection centers is $L_s=2$ µm along the straight and $L_b=3.2$ µm along the bent waveguide. The waveguides encircle an area of about $A=1.63$ µm$^2$. In the following we label the four leads as 1, 2, 3, and 4, according to Fig. 1 (a). The leads are used to separate current injection and extraction from voltage probing. In particular, interchanging the contacts allows us to investigate asymmetries in the two paths of the quantum ring structures. In Fig. 1 (b) the local four-terminal measurement setup is shown in contrast to the non-local measurement as in Fig. 1 (c). Transport measurements were performed at $T$=1.5 K.

With a semiconductor parameter analyzer DC characteristics are measured in a push-pull fashion, in which the input voltage $V_{ij}=V_i-V_j$ with $V_j=-V_i$ is applied to the probe pair $(i,j)$. Simultaneously the voltage drop $V_{kl}=V_k-V_l$ is detected and the resulting input current $I_{ij}$ is recorded. Fig. 2 (a) shows the remote bend resistance characteristic of the four-terminal ring structure at fixed gate voltages and zero magnetic field. For small input currents $|I_{12}|<2$ µA a strong negative non-local resistance $R_{34,12}=V_{34}/I_{12}$ is measured. This is a clear signature for ballistic transport as commonly reported.[17,18] The negative non-local resistance $R_{34,12}$ results from the collimation results from the collimation effect of inertial ballistic electrons. In the semi-classical billard picture,[19] for $I_{12}<0$ ($V_{12}<0$ applied) electrons injected from lead 1 are efficiently transmitted into lead 4 along the straight waveguide inducing a positive voltage $V_{34}$. The negative resistance $R_{34,12}$ is a signature for the collimation efficiency. For positive $I_{12}>0$ electrons are collimated into lead 3 along the *bent* waveguide. The same slope in the V-I-characteristic for negative and positive $I_{12}$ ($R_{34,12}(I_{12}<0) = R_{34,12}(I_{12}>0)\approx-260$ Ohm) reveals that the ballistic injection along the longer bent waveguide is as efficient as the injection along the shorter straight waveguide. Such a behavior has not been reported before, however, it can be expected due the symmetric orthogonal cross junctions.

We observe local extrema in $V_{34}$ in the V-I characteristic indicating the cross-over from ballistic to diffusive transport by carrier heating ($|I_{12}|=3.7\pm0.3$ µA which corresponds to an excitation voltage of 9.8 mV, Fig. 2(a)). These result from electron-electron scattering at a certain excess energy.[20] For higher input currents the non-local voltage $V_{34}$ saturates due to an increased scattering of hot electrons with optical phonon emission.[21]

The V-I-characteristic in the cross-local setup is shown in Fig. 2 (c). Here, the driving push-pull voltage was applied between leads 1 and 3, and the voltage drop between leads 2 and 4 was recorded. For high input currents the resistance is positive, while it is almost zero for $-1$ µA$<I_{13}<1$ µA. The positive resistance clearly shows the diffusive nature of the transport, where electrons follow the applied electric field. For small input currents, the transport has a strong inertial ballistic component which induces a negative $R_{24,13}$ and competes with the diffusive component. Both components cancel each other for both positive and negative currents of up to 1µA. This is also a signature for the same ballistic injection efficiency along the bent and the straight waveguides.

The magnetic field dependence of the non-local voltage for an AC excitation current $I_{12,ac}$ of 50 nA is shown in Fig. 2 (b). As required for the AB-measurements discussed below, the magnetic field is applied perpendicular to the plane of the quantum wire ring. Around zero magnetic field, $V_{34}$ is negative. It approaches zero for higher magnetic fields ($|B|\sim250$ mT) at which the Lorentz force changes the trajectories of the ballistic electrons substantially and suppresses the negative non-local bend resistance. However, for magnetic fields $|B|\sim<100$ mT the collimation effect along both the bent and the straight waveguides is still efficient.

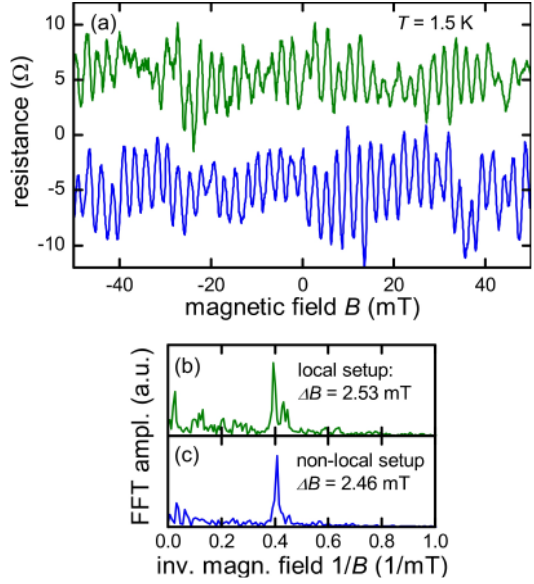

Fig. 4. (a) AB-component of $R_{23,14}$ and $R_{34,21}$ after substraction of the aperiodic background (both offset by 5 Ohms for clarity). FFT power spectrum of the (b) local and (c) non-local AB-oscillations shown in (a).

For quantum interference experiments we measure the differential resistance $(dI/dV)^{-1}$ via standard lock-in technique with AC currents of $\leq 50$ nA at a frequency of 133 Hz. The magnetic field of an electromagnet is varied quasi-statically and the measurement data recorded with minimal step width of 10 µT or less. The magnetoresistance traces for the local and non-local setup are shown Fig. 3 (a) and 3 (b), respectively. In both configurations, the AC driving current was 28 nA at a frequency of 133 Hz. The measurements are plotted with a moving average of 15 data points. Periodic AB-conductance oscillations are superimposed on an aperiodic resistance variation which can be attributed to universal conductance fluctuations. To determine the period of the AB oscillations we subtracted the aperiodic background with a moving average of 200 data points (Fig. 4 (a)) and performed fast Fourier transforms (FFT) shown in Fig. 4 (b,c). As a result, the obtained values for the AB conductance oscillation period are 2.53±0.05 mT for the local and 2.46±0.05 mT for the non-local configuration, and which are therefore nearly equal within the experimental errors. They are also in good agreement with the expected value of $\Delta B = h/(eA) \sim 2.55 \pm 0.3$ mT from the geometrical estimate.

In the Landauer-Büttiker (LB) formalism for phase coherent conductors, the four-terminal resistance is given by $R_{mn,kl} = (h/Ne^2)(T_{km}T_{ln}-T_{kn}T_{lm})/D$, where $T_{ij}$ is the transmission coefficient for electrons form terminal $j$ to terminal $i$,[22] and $N$ is the number of transmitting modes. $D$ is the subdeterminant of the transmission matrix. The four-terminal geometry (Fig. 1) can be reduced to three different transmission probabilities $T_0$, $T_1$, and $T_2$: $T_0 \equiv T_{12} = T_{21} = T_{34} = T_{43}$, $T_1 \equiv T_{14} = T_{41}$, and $T_2 \equiv T_{24} = T_{42} = T_{13} = T_{31}$.

Furthermore, from the symmetry of the orthogonal cross junctions we expect that $T_{23} = T_{14}$, so that $T_1 \equiv T_{14} = T_{41} = T_{23} = T_{32}$. Electron interference and decoherence when transversing the ring can be taken into account by $T_{ij} = \alpha_{ij} + \beta_{ij} * \cos(2\pi\Phi/\Phi_0+\varphi)$, where $\alpha_{ij}$ is the regular transmission superimposed with the Aharonov-Bohm part $\beta_{ij}$. Here, $\Phi$ is the magnetic flux encircled, $\Phi_0 = h/e$ is the flux quantum, and $\varphi$ is a sample specific phase. With the above approximation, the LB formalism yields to the subdeterminant of $D = (T_0+T_1)(T_1+T_2)(T_0+T_2)$ and gives

$$R_{23,14} = (h/Ne^2)(T_0-T_2) / [(T_0+T_1)(T_1+T_2)]$$
$$R_{34,12} = (h/Ne^2)(T_1-T_2) / [(T_0+T_1)(T_0+T_2)] \quad (1)$$
$$R_{24,13} = (h/Ne^2)(T_0-T_1) / [(T_1+T_2)(T_0+T_2)]$$

for the local, the non-local and the cross-local setup, respectively. Taking into account that the local and non-local resistances are nearly equal $R_{23,14} \sim R_{34,21}$ (Figs. 3 (a) and (b)), we conclude from Eq. (1) that $T_0 \sim T_1$. This is in good agreement with the zero resistance regime observed in the cross-local measurement in Fig. 2 (c).

In summary, for the asymmetric quantum ring structure with orthogonal cross-junctions we showed: On one hand, the conductance behaviour is clearly ballistic ($T_0 \sim T_1$). On the other hand, we find significant AB conductance oscillations with a high relative amplitude at temperatures of more than $T \sim 1.5$ K. This demonstrates coherent quantum transport in both paths of the asymmetric quantum ring. In contrast to rings with non-orthogonal lead cross junctions[15] the amplitudes are *equal* for the local and the non-local setup. We conclude, firstly, that a strong asymmetry in the ring does not hamper the detection of quantum interference and, secondly, that tailoring of the detailed geometry of the cross junctions can be used to enhance the non-local detection of the AB amplitude significantly. This is an important step towards usage of quantum intereference devices for implementation of interferometers based on quantum wires in nanoelectronics or spintronics.

Our work was supported partly by the German Federal Ministry of Education and Research (BMBF) and the German Science Foundation (DFG-SPP 1285, FI-932-1).

[1] Y. Aharonov, D. Bohm, Phys. Rev. **115**, 485 (1959).
[2] Y. Imry, R. A. Webb, Scientific American **260**, 56 (1989).
[3] A. Fuhrer, *et al.*, Nature **413**, 822 (2001).
[4] R. Schuster, *et al.*, Nature **385**, 417 (1997).
[5] K. Kobayashi, *et al.*, Phys. Rev. Lett. **88**, 256806 (2002).
[6] M. Sigrist, *et al.*, Phys. Rev. Lett. **96**, 036804 (2006).
[7] M. Knop, *et al.,* Appl. Phys. Lett. **88**, 082110 (2006).
[8] D. Schmeltzer, *et al.*, Phys. Rev. B **68**, 195317 (2003).
[9] R. Ionicioiu, I. D´Amico, Phys. Rev. B **67**, 041307 (2003).
[10] T. Koga, *et al.*, Phys. Rev B **70**, 161302 (2004).
[11] U. Zülicke, Appl. Phys. Lett. **85**, 2616 (2004).
[12] P. Devillard, *et al.*, Europhys. Lett. **74**, 679-685 (2006).
[13] c.f. A.E. Hansen, *et al.*, Phys. Rev. B **64**, 045327 (2001).
[14] c.f. M. Cassé, *et al.*, Phys. Rev. B **62**, 2624 (2000).
[15] K. Kobayashi, *et al.,* J. Phys. Soc. Jap. **71**, 2094 (2002).
[16] G. Seelig, *et al.*, Phys. Rev. B **68**, 161310 (2003).
[17] G. Timp, *et al.*, Phys. Rev. Lett. **60**, 2081 (1988).
[18] Y. Takagaki, *et al.*, Solid State Commun. **68**, 1051 (1988).
[19] C.W.J. Beenakker, H. van Houten, Phys. Rev. Lett. **63**, 1857 (1989).
[20] D.R.S. Cumming, J.H. Davies, Appl. Phys. Lett. **69**, 3363 (1996).
[21] U. Sivan, *et al.*, Phys. Rev. Lett. **63**, 992 (1989).
[22] M. Büttiker, Phys. Rev. Lett. **57**, 1761 (1986).